\documentclass[modern]{aastex61}

\newcommand\nimg{$\approx 3000$}
\newcommand\ntrain{2004}  
\newcommand\nnormtrain{3885} 
\newcommand\ntotaltrain{15540} %
\newcommand\nvalid{498}
\newcommand\ntotalvalid{3880}
\newcommand\ntest{626}
\newcommand\ntestnorm{4840}
\newcommand\narc{121}
\newcommand\nscience{595}
\newcommand\nstandard{181}
\newcommand\nbias{330}
\newcommand\nflat{777}
\newcommand\spit{{\sc SPIT}}
\newcommand\pypit{{\sc PYPIT}}
\newcommand\perctrain{98.7\%} 
\newcommand\perctest{98.7\%}  
\newcommand\percftest{98.2\%}  

\received{May 3, 2018}
\accepted{June 16, 2018}
\submitjournal{PASP}

\shorttitle{The SPectral Image Typer (\spit)}
\shortauthors{Jankov \& Prochaska}

\begin{document}

\title{Spectral Image Classification with Deep Learning}

\correspondingauthor{J. Xavier Prochaska}
\email{xavier@ucolick.org}

\author{Viktor Jankov}
\affil{Computer Science \\ 
UC Santa Cruz, 1156 High St. \\
Santa Cruz, CA 95064 USA} 

\author{J. Xavier Prochaska}
\affil{Astronomy \& Astrophysics \\
UC Santa Cruz, 1156 High St. \\
Santa Cruz, CA 95064 USA 
}



\begin{abstract}

We present the Spectral Image Typer (\spit), a convolutional neural
network (CNN) built to classify spectral images.
In contrast to traditional, rules-based algorithms which rely
on meta data provided with the image (e.g.\ header cards),
\spit\ is trained solely on the image data.  We have trained \spit\
on \ntrain\ human-classified images taken with the Kast spectrometer
at Lick Observatory with types of Bias, Arc, Flat, Science and Standard. 
We include several pre-processing steps (scaling, trimming) motivated
by human practice and also
expanded the training set to balance between
image type and increase diversity.
The algorithm achieved an accuracy of \perctrain\ on
the held-out validation set and an accuracy of
\perctest\ on the test set of images.
We then adopt a slightly modified classification scheme to
improve robustness at a modestly reduced cost in 
accuracy (\percftest).
The majority of mis-classifications are Science frames
with very faint sources confused with Arc images
(e.g.\ faint emission line galaxies) or Science
frames with very bright sources confused with Standard
stars.  These are errors that even a 
well-trained human is prone to make.
Future work will increase the training set from Kast,
will include additional optical and near-IR instruments,
and may expand the CNN architecture complexity.
We are now incorporating \spit\ in the \pypit\ data reduction
pipeline (DRP) and are willing to facilitate 
its inclusion in other DRPs.

\end{abstract}




\section{Introduction} \label{sec:intro}

The construction of an astronomical spectrum from CCD images
acquired on modern facilities is a complex procedure of image
processing, manipulation, extraction, and combination
\citep[e.g.][]{bernstein+15}.  This includes corrections for detector
electronics and imperfections, the conversion of detector
geometry to spectral (e.g.\ wavelength) and spatial dimensions, 
and the calibration of detector counts to physical flux units.
To perform these steps, one generally obtains
a series of calibration images that complement the on-sky,
science exposures.  One must then organize and associate
the various types of spectral images with the
science frames (prior to processing) before continuing
with data reduction. 

The first step in a DRP, therefore,
is to identify the `type' of each spectral image.  
The common set of spectral image types includes:

\begin{itemize}
\item Bias -- a short exposure (ideally zero seconds) used to characterize
  the bias count level associated with the detector electronics.
\item Dark -- an exposure of finite duration, typically matched to the
  exposure time of the science frame, taken without any
  input light.  This characterizes the dark current of the detector and
  electronics.
\item Arc -- a spectral image of one or more calibration arc lamps (e.g. ThAr).
\item Flat -- a spectral image of a continuum light source (e.g. an incandescent 
  lamp).
\item Standard -- a spectral image of a bright star with a previously measured flux
  at the wavelengths of interest.
\item Science -- a spectral image of one or more science targets.
\end{itemize}

For most instruments and observatories, meta data describing the image, 
instrument, and telescope are saved together with the spectral image.
In the Flexible Image Transport System (FITS) format,
the most common format in astronomy,  an ASCII Header
provides this information in a series of value/comment `cards'.
From this Header, one may develop a rules-based algorithm to 
decipher the meta data\footnote{Some instruments even dedicate a Header
card to designate the image type.} 
and thereby assign a type to the spectral image.
In practice, however, this Header-based process of image-typing 
suffers from
 (i) insufficient or ambiguous information;
 (ii) poor human behavior (e.g.\ leaving the CCD shutter open when taking a dark frame);
 (iii) unanticipated failures of the instrument, telescope, software, or hardware
 (e.g. a failed arc lamp).
These issues complicate image-typing and lead to erroneous processing
and data reduction.
A conscientious observer, therefore, tends to visually inspect each
image for quality assurance and to correct the image type as needed.
This laborious step is frequently performed with an image viewer and the results
are manually recorded to a file (another source of human error).

Another flaw in the standard process is the specificity of rule-based algorithms.
While spectral images of the same type often look very similar across instruments, 
there is great diversity amongst their Header data.  This includes 
a non-uniform set of cards and values.
The practice is sufficiently diverse that one cannot adopt a single code
to handle all instruments (or even a pair of instruments!).
Instead, one must develop a new algorithm (or parameters within that
algorithm) and then test it extensively.
This practice is time consuming and difficult to maintain and update.

Recognizing the challenges and loss of productivity that image-typing
imposes, we have taken a deep learning approach to
the problem.  Recent developments in image classification with
convolutional neural networks \citep[CNNs][]{lecun-89t} and our own
success with classifying features in one-dimensional
spectra \citep{parks+18}, inspired
our development of a CNN algorithm for spectral image-typing.
In this manuscript, we report on our first results with the
Spectral Image Typer (\spit), release a product
that will be expanded to include instruments across the globe,
and describe the prospects for future development.

This paper is organized as follows:
Section~\ref{sec:training} presents our image training set,
the CNN is described in Section~\ref{sec:cnn},
Section~\ref{sec:validation} reports on the validation results,
and Section~\ref{sec:future} describes future work and development.
The \spit\ repository is available on GitHub as part
of the \pypit\ organization:  https://github.com/PYPIT/spit.

\section{Training Set} 
\label{sec:training}

\begin{figure}
\begin{center}
\includegraphics[width=5.5in]{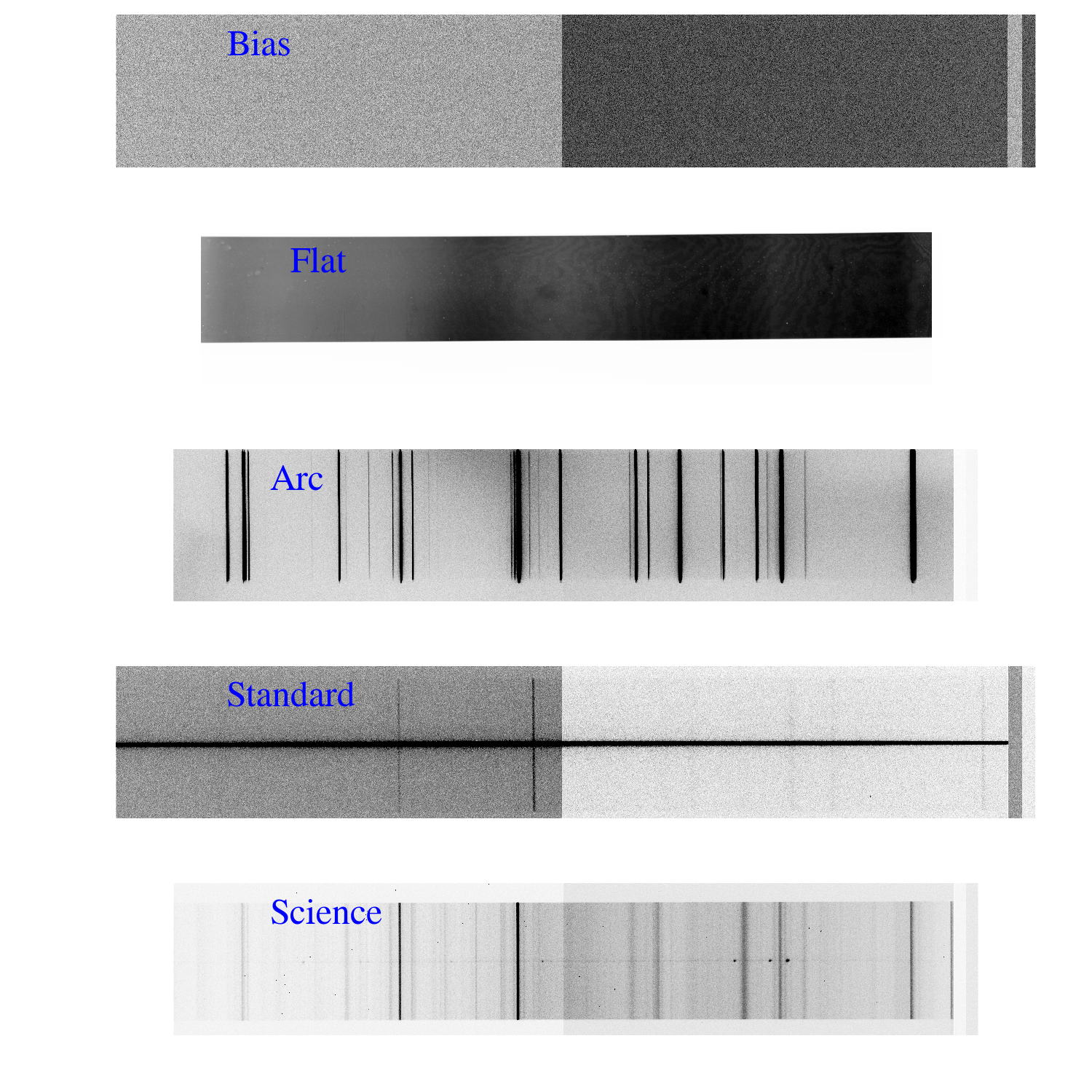}
\caption{
Example set of spectral images
from the blue and red cameras of the Kast spectrometer
labeled by image type.  The images have been 
pre-processed with the {\sc zscale} algorithm to
emphasize spectral features.
}
\label{fig:images}
\end{center}
\end{figure}

In many respects, a CNN is only as good as its training set.
For this first effort, we focus on a single instrument
-- the Kast spectrometer\footnote{https://mthamilton.ucolick.org/techdocs/instruments/kast/} 
on the 3m Shane telescope of Lick Observatory.
The Kast instrument is a dual camera spectrometer with one
camera optimized for each of {\it blue} and {\it red} wavelengths.
Kast was deployed at Lick Observatory in February 1992
and has been scheduled for nearly 5,000 nights on the Shane 3m.
Figure~\ref{fig:images} shows a fiducial set of spectral images
from this instrument, labeled by image type.  
Co-author JXP and his collaborators have obtained thousands of spectral 
images with the Kast over the past several years from which
we have derived the training set detailed below.

\subsection{Data}

We collated \nimg\ images obtained on a series of observing runs of 
PI Prochaska and PI Hsyu 
from 2008 to 2016.  Many of these images were typed by 
rules-based algorithms of the 
\href{http://www.ucolick.org/~xavier/LowRedux/}{LowRedux}
or
\href{http://pypit.readthedocs.io/en/latest/}{PYPIT}
DRPs and then verified by visual inspection.
The latter effort was especially necessary 
because the Header of the 
blue camera was erroneous during a significant interval
in this time period (e.g.\ incorrect specification of the calibration
lamps that were off/on).
From this full dataset, we selected \ntrain\ images for training,
which separate into \nbias~Bias images,
\narc~Arc frames, \nflat~Flat frames, 
\nscience~Science frames, and \nstandard~Standard images.
We have not included any Dark frames in this analysis\footnote{
We suspect the algorithm would struggle to distinguish Dark frames from Bias frames for optical detectors (although it could learn the incidence of cosmic rays).}.

\begin{deluxetable}{lccccc}
\tablewidth{0pc}
\tablecaption{Training Set\label{tab:images}}
\tabletypesize{\small}
\tablehead{\colhead{Type?} & \colhead{Date} 
& \colhead{Frame} 
& \colhead{Usage} 
} 
\startdata 
arc & 2008-09-25 & b10004 & train\\ 
arc & 2008-09-25 & b10006 & test\\ 
arc & 2008-09-25 & r10004 & train\\ 
arc & 2008-09-26 & b1005 & train\\ 
arc & 2008-09-26 & b1007 & train\\ 
arc & 2008-09-26 & r1005 & test\\ 
arc & 2008-09-27 & b2007 & train\\ 
arc & 2008-09-27 & b2009 & test\\ 
arc & 2008-09-27 & r2007 & test\\ 
arc & 2008-09-28 & b3001 & train\\ 
arc & 2008-09-28 & r3001 & train\\ 
arc & 2008-09-29 & b4002 & test\\ 
arc & 2008-09-29 & r4005 & validation\\ 
arc & 2010-11-05 & b3000 & validation\\ 
arc & 2010-11-05 & b3001 & validation\\ 
arc & 2010-11-05 & b3002 & train\\ 
\enddata 
\end{deluxetable}

Table~\ref{tab:images} lists all of the images and the assigned image type.
This dataset is the starting point for the training set of our CNN.  
From these images,
we held-out \nvalid\ images for validation of the algorithm 
after each training iteration, as listed 
in the Table, and we also held-out
\ntest\ images for testing 
None of the validation nor test
files were included in training.
Each of the Kast spectral images is stored as 
an unsigned integer (16-bit) array in an individual
FITS file. 
Every FITS file contains a Header, but these were
entirely ignored in what follows.

Conveniently, there is significant variability within
each class in the training set.  The Arc frames, for example,  
were obtained with several grisms and gratings
covering different spectral regions and with differing spectral
resolutions.  Furthermore, there was variation in the arc lamps
employed by different observers and the exposure times adopted.
Similarly, the Flat frames of the two cameras have considerably
different count levels on different portions of the CCD.
This variability will help force the CNN to `learn' low-level features
of the various image types.

Most of the spectral images obtained with Kast during this period 
used detectors that have the spectral dimension oriented along
the detector rows, i.e.\ wavelengths lie along NAXIS1 in the FITS Header nomenclature
and NAXIS1~$>$~NAXIS2.
The few that were oriented with the spectral axis parallel
to NAXIS2 were rotated 90~degrees prior to processing.
As described below, we flip each image in the horizontal and
vertical dimensions to increase
the size and variability of the training set.

\subsection{Pre-processing}

Prior to training, each input image was processed in
a series of steps.  Two of these steps, not
coincidentally, are often 
performed by the human during her/his own image typing:
one explict and one implicit.  
The {\it implicit} pre-processing step performed by a human is
to ignore any large swaths of the image that
are blank.  Such regions result when the slit 
is shorter than the detector or other instrument vignetting
occurs.  One may also have an extensive overscan region,
i.e.\ comparable to the size of the data region.
This overscan is included in all of the spectral images,
independent of type.
Initially, large blank regions confused the CNN
algorithm into classifying images as a Bias and
we introduced a trimming pre-processing
step to address this failure mode.

\begin{figure}
\begin{center}
\includegraphics[width=3.5in]{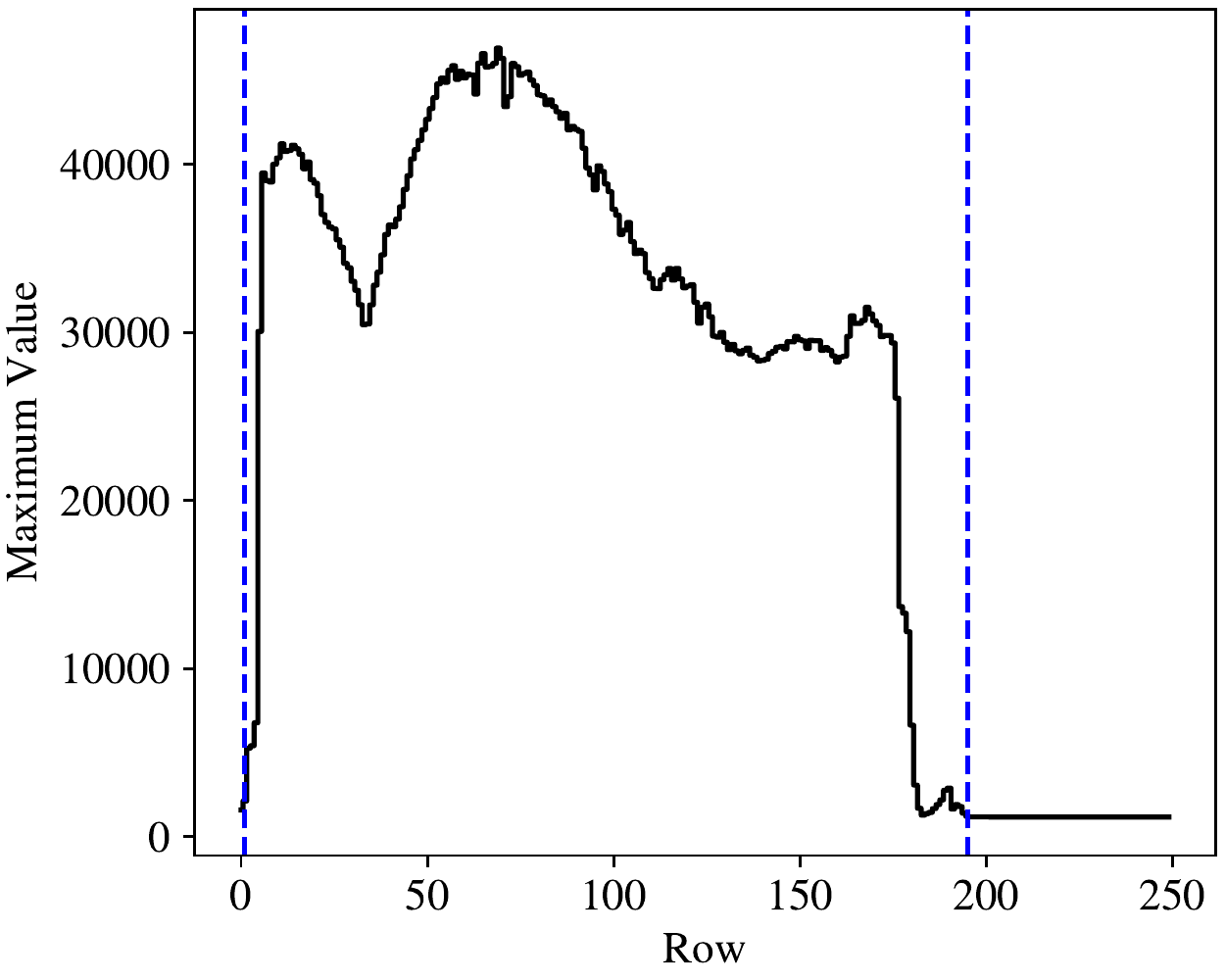}
\caption{
Plot of the maximum values in each row of an Arc image
after performing a {\sc sigma\_clip} iteration.
The blue dashed lines show the starting and end rows
of the data region defined by our pre-processing step
(see the text for further details).
}
\label{fig:find_trim}
\end{center}
\end{figure}

\begin{figure}
\begin{center}
\includegraphics[width=5.5in]{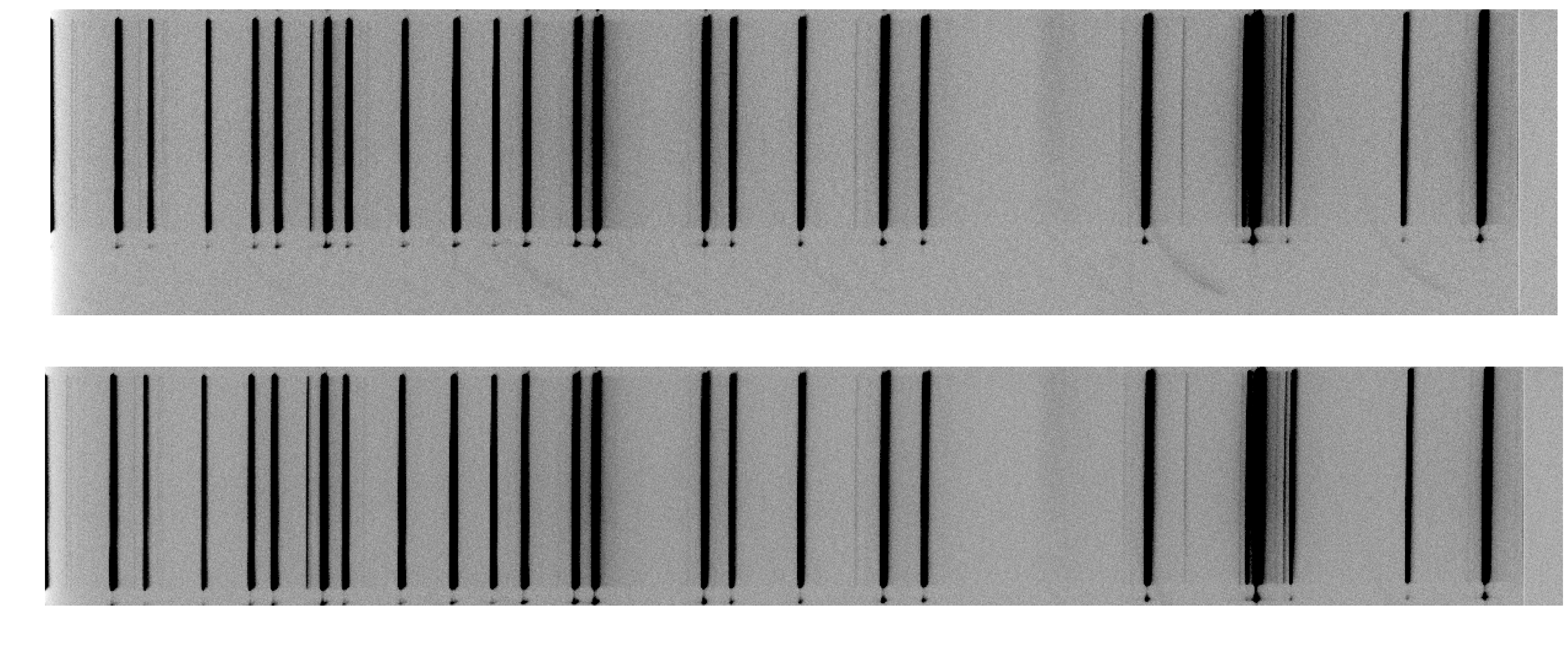}
\caption{
The top panel shows an Arc frame that has been scaled
with {\sc zscale} but is otherwise original and includes
a significant, blank region in the lower 1/4 of the image.
The bottom panel shows the same spectral image after
trimming the empty regions from each side.
}
\label{fig:trim}
\end{center}
\end{figure}

We identified the data regions of our spectral images as follows.
Figure~\ref{fig:find_trim} shows the
maximum value of each row in an example Arc image
(specifically frame {\it r6} from 18 Aug 2015).
To analyze the image we:
(i) applied the {\sc astropy} sigma\_clip algorithm
on the image 
along its zero axis (i.e.\ down each row).  
This removes any extreme outliers (e.g.\ cosmic ray events);
(ii)  iterated through each row and recorded the maximum
values as displayed in the figure;
(iii)  iterated over the maximum value array twice, once from the front, and another time
from the back. For each index value, we
check if the current value is greater than the previous value
by more than 10\%.
Bias or blank regions have count levels stable to 10\%\ and 
are ignored;  any large departures indicate the data region.
Figure~\ref{fig:find_trim} indicates the region determined in this 
fashion and 
Figure~\ref{fig:trim} shows the Arc frame before and after trimming.

The next step in the preprocessing pipeline are to 
lower the image resolution. 
The input images are unnecessarily large
and  would have made the computation excessively long. 
Therefore, each image was resized to 210x650 pixels (spatial, spectral) using linear interpolation\footnote{Performed with 
the {\sc congrid} task provided
in the {\sc scipy} documentation.}.

\begin{figure}
\begin{center}
\includegraphics[width=5.5in]{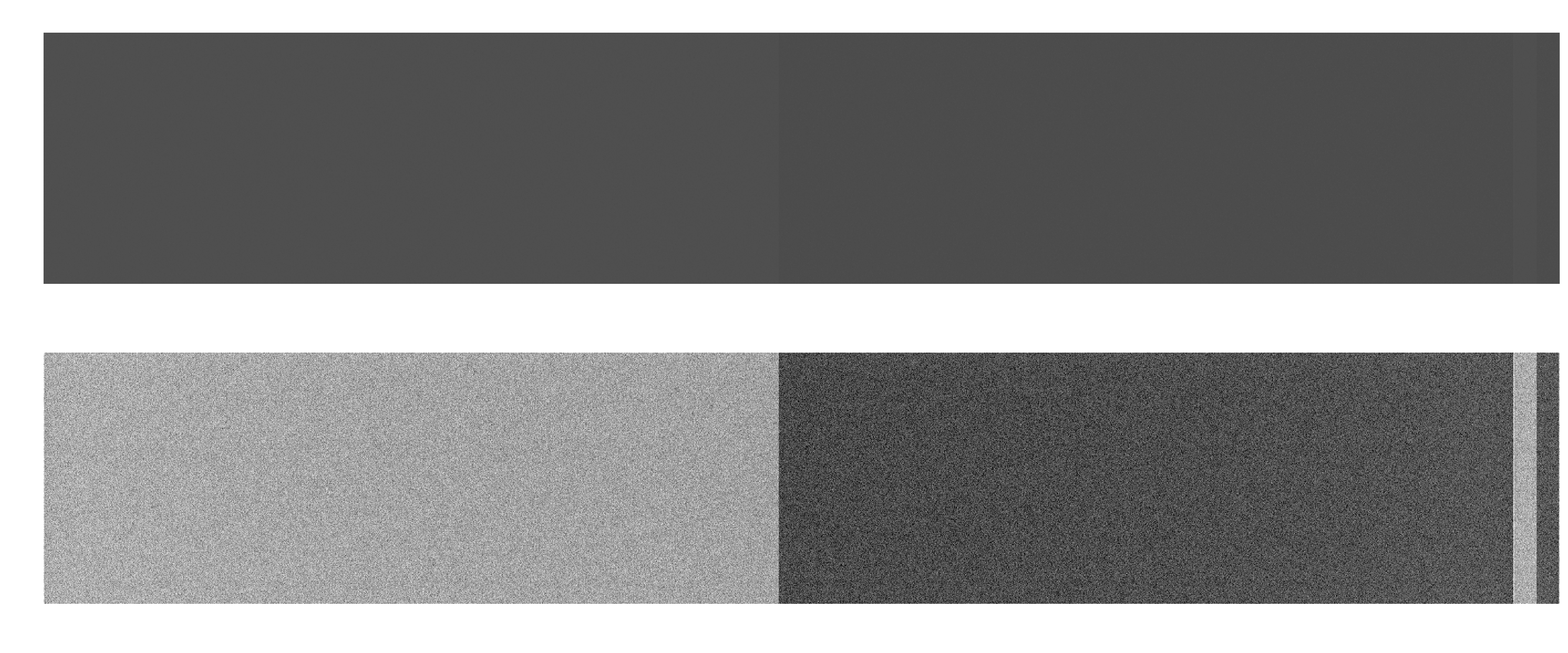}
\caption{
Example Bias image visualized before (top) and after (bottom)
applying the {\sc zscale} algorithm.
}
\label{fig:zscale}
\end{center}
\end{figure}

The {\it explicit} step taken by a human is to visualize
the data with an image viewer that
sets the color-map to a finite range of the flux. 
A popular choice to perform this rescaling
is the {\sc zscale} algorithm which is designed
to set the color-map to span a small interval
near the median value of the image. 
This algorithm has been applied to adjust the color-map limits 
for the images in Figure~\ref{fig:images}, and 
views with and without the scaling is shown
for a Bias image in Figure~\ref{fig:zscale}.

Briefly, the {\sc zscale} algorithm generates an 
interval $[z_1, z_2]$ for
an image $I(i)$ with $n_{\rm pix}$ flux values as:

\begin{eqnarray}
z_1 = I(i_{\rm mid}) + m (1 - i_{\rm mid})/C \\
z_2 = I(i_{\rm mid}) + m (n_{\rm pix} - i_{\rm mid})/C  \;\;\; ,
\end{eqnarray}
where $i_{\rm mid}$ is the central pixel of the sorted image array,
$C$ is the contrast (set to 0.25), 
and the slope $m$ is determined by a linear function fit with iterative rejection,

\begin{equation}
I(i) = b + m * (i - i_{\rm mid})
\end{equation}
If more than half of the points are rejected during fitting, then there 
is no well-defined slope 
and the full range of the sample defines $z_1$ and $z_2$. 
Finally, the image was scaled to values ranging
from $0-255$ and recast as unsigned, 8-byte integers.
The image may then be saved as a PNG file to 
disk, without the Header.

\subsection{Normalization and Diversification}

Out of the \ntrain\ images in the original dataset, there was an imbalance
in the various image types owing to the typical approach of astronomers to
acquiring data and calibrations. 
This was a major issue because some of the more important classes, which are harder to
distinguish from each other (e.g. Arcs from Sciences), were significantly underrepresented. 
Therefore, for the training and validation sets
we oversampled the under-represented types and
made the image count equal between all image types.  
This meant including additional, identical images within 
each of the types except Flat frames.
As an example,  we replicated each of the \narc\ Arc images
six times for 756~images and then drew an
additional 21 Arc images at random to match the \nflat\ Flat frames.
This yielded \nnormtrain\ images across all types.

We further expanded the training set by creating three additional
images by flipping the original along
each axis: horizontal, vertical, and horizontal-vertical.  
This yielded four images per frame, including the original.
This pre-processing step
helps the CNN learn more effectively by forcing 
a more flexible algorithm and one that is forced to learn 
patterns instead of simply memorizing the set of 
images. 
Further work might consider  randomly cropped images, or random 
changes to the hue, contrast and saturation. 

Altogether, we had \ntotaltrain\ 
training set images that have also been scaled by 
{\sc zscale},  trimmed to remove the overscan and any other blank regions,
and resized to 210x650 pixels. 
Normalization and diversification (i.e.\ flipping) was also
performed on the set of validation images yielding a total
\ntotalvalid\ images that are classified during validation.

\section{The CNN}
\label{sec:cnn}

\begin{figure}
\begin{center}
\includegraphics[width=5.5in]{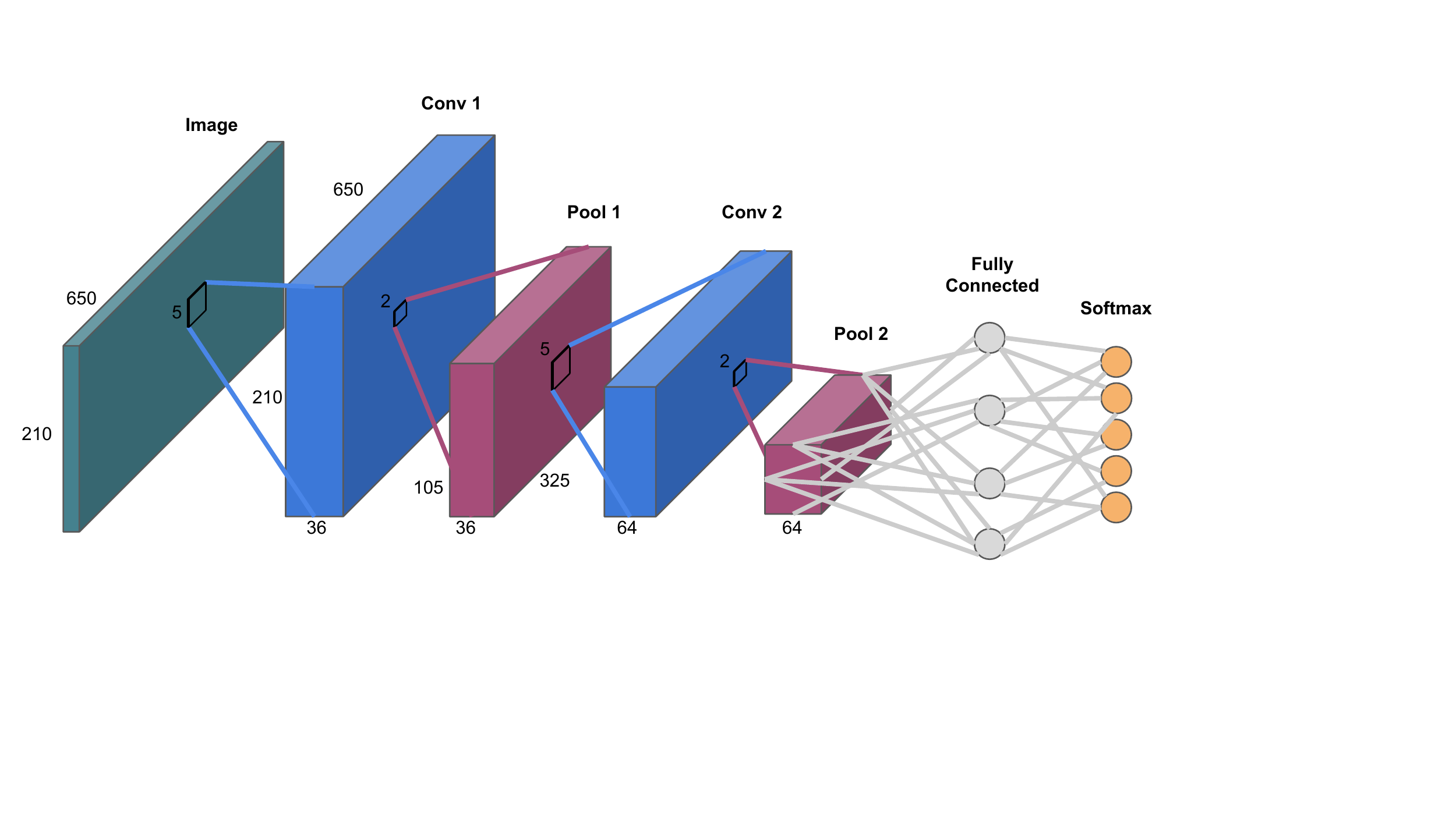}
\caption{Schematic of the CNN architecture that underlies
\spit.  We implement two convolutional layers, each with a stride
of 1, two maxpool layers each with a stride of 2, and end with 
a fully connected layer and a softmax classifier.
}
\label{fig:arch}
\end{center}
\end{figure}

\subsection{Architecture}

To better describe the CNN architecture, we first define a few terms 
that relate to CNNs in general:

Kernel -- 
CNNs work with filter scanners, which are NxM matrices that
scan the input image and accentuate specific patterns. 
For our architecture, we adopted the standard 5x5 kernel that is frequently
used in image processing CNNs (e.g.\ Inception).

Stride -- 
Stride refers to the step-size for moving
the kernel across the input image.  A stride of 1 means the kernel 
is moved by 1 pixel in the
horizontal direction, and at the end of the row it
is moved 1 pixel down to continue scanning. 

Depth -- Depth describes the number of features that the CNN aims to learn by
scanning the image.  
The larger the depth, the more complex features the network may
learn and the more complex patterns that can be classified. 

Max\_pool -- Max pooling refers to down sampling the original image to prevent
overfitting while also enhancing the most important features.  
This is achieved by reducing the dimensionality of the input image. 
Max pooling also provides an
abstracted form of the original representation. 
Furthermore, it reduces the computational cost by
reducing the number of parameters to learn.
Our architecture applies max\_pool layers of 2x2 size.

Softmax classifier -- Our classifier uses the softmax function which creates a
probability distribution over K different outcomes. In our CNN, the softmax classifier 
predicts the probability of an image with respect to the five different classes. For example, 
an input Bias image might produce an output of [0.8, 0.1, 0.1, 0, 0],
which means that the network has 80\%\ confidence that the image
is a Bias image, and 10\%\ that it is a Science frame or a
Standard frame. In a softmax function, the values range from 
0 to 1 and sum to unity.

Convolutional Layer -- A convolutional layer  
connects neurons to local regions in the input image.  
These layers are frequently the initial set in a CNN. 

Fully Connected (FC) Layer -- 
Unlike a Convolutional layer, the FC 
has all of the neurons connected to the entire previous layer. 
Typical practice is to use an FC layer at
the end of the architecture.

One-hot encoded array -- This array describes
the original classes as a simple matrix. 
For a Bias image,
the CNN represents the frame as [1, 0, 0, 0, 0], i.e.\  the Bias class
corresponds to the first index.  The complete one-hot encoding
was done as follows: 
0 is Bias, 1 is Science, 2 is Standard, 3 is Arc, and 4 is a Flat frame.

\vskip 0.1in

The CNN we developed has the following architecture. 
It begins with a Convolutional
Layer with a 5x5 kernel, a stride of 1 pixel, and a depth of 36. It is followed 
by a max\_pool layer of 2x2, 
followed by another convolutional layer with a 5x5 kernel, a 1 pixel stride,
and a depth of 64.  This layer is also followed by a max\_pool 
layer of 2x2.   The architecture finishes with a fully 
connected layer of 128 features with a softmax classifier at the end 
to produce a one-hot encoded array. 
Figure~\ref{fig:arch} summarizes the architecture.
Several, slightly modified architectures were also explored
but the adopted model offered excellent performance
with relatively cheap execution\footnote{A strong constraint for this time-limited 
Master's thesis in a GPU-limited environment.}.

\subsection{Training}

For training, we used the Adam 
\citep[Adaptive Moment Estimation][]{Adam2015}
bundled with Google's deep learning {\sc Tensorflow} \citep{tensorflow2015-whitepaper}
with a learning rate of $10^{-4}$
and no dropout rate.
Each epoch trained on a random batch of 30 images and
then was assessed against the validation image set.
We ran for a series of epochs
on a NVIDIA Tesla P100-SXM2 GPU, taking approximately 6s per update.
After $\approx 1,000$ epochs, we noticed no significant learning.
The highest accuracy achieved on the validation set during
training was \perctrain,  and this is the model 
saved for classification.


\section{Results}
\label{sec:validation}

Table~\ref{tab:images} lists the \ntest\
images held-out for testing the accuracy
of the optimized CNN classifier.   
The set is comprised of 
39~Arc, 
103~Bias,
242~Flat,
186~Science,
and 56~Standard images from the blue and
red cameras of Kast.
We performed several tests and assessed
the overall performance of the CNN as follows.

\begin{figure}
\begin{center}
\includegraphics[width=5.5in]{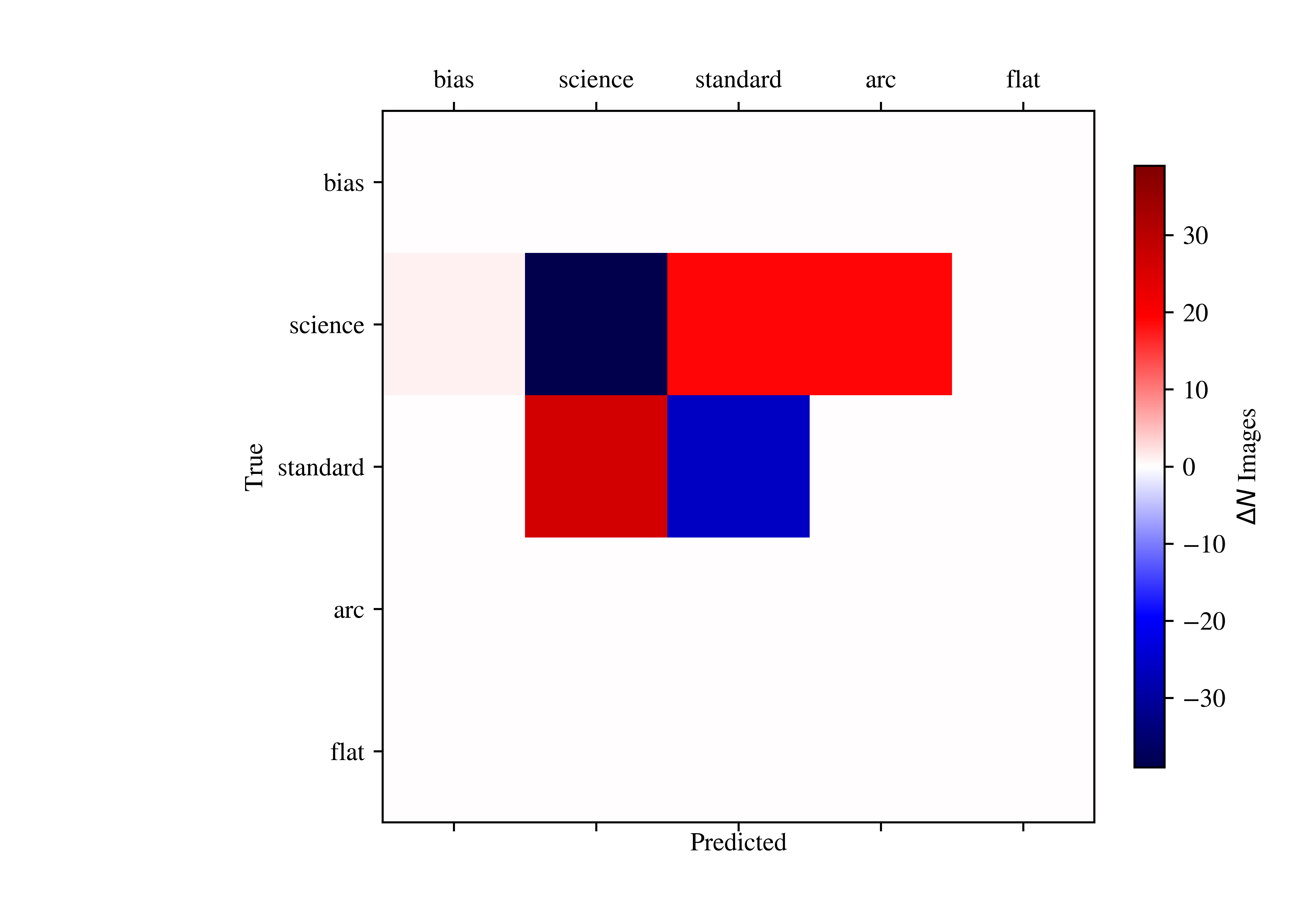}
\caption{Differential $\Delta N$ confusion matrix describing the accuracy
of the optimized CNN classifier on the normalized test suite.
On the diagonal, negative values indicate the number of images
that were misclassified (out of a total of 968).  
Off the diagonal, the positive values
indicate the confusion between image types.
A perfect classifier would present a differential confusion
matrix of all zeros.
}
\label{fig:test_norm}
\end{center}
\end{figure}

\subsection{Normalized Accuracy}

We first report results by maintaining the same
methodology taken for generating the training
and validation sets.  Specifically, we modified
the test set to have identical numbers of each
image type and then assessed each of these images
flipped along its horizontal and vertical axes
(4 orientations for each image).  Each of these
was classified according to the majority value
of the one-hot encoded array produced by the 
optimized CNN classifier.

Figure~\ref{fig:test_norm} shows the differential
confusion matrix for this normalized test set of 
\ntestnorm\ images.  The test accuracy is \perctest,
and all of the confusion occurs for the Science
and Standard frames.  These results are consistent
(nearly identical) to those recorded for the validation
set.

\begin{figure}
\begin{center}
\includegraphics[width=5.5in]{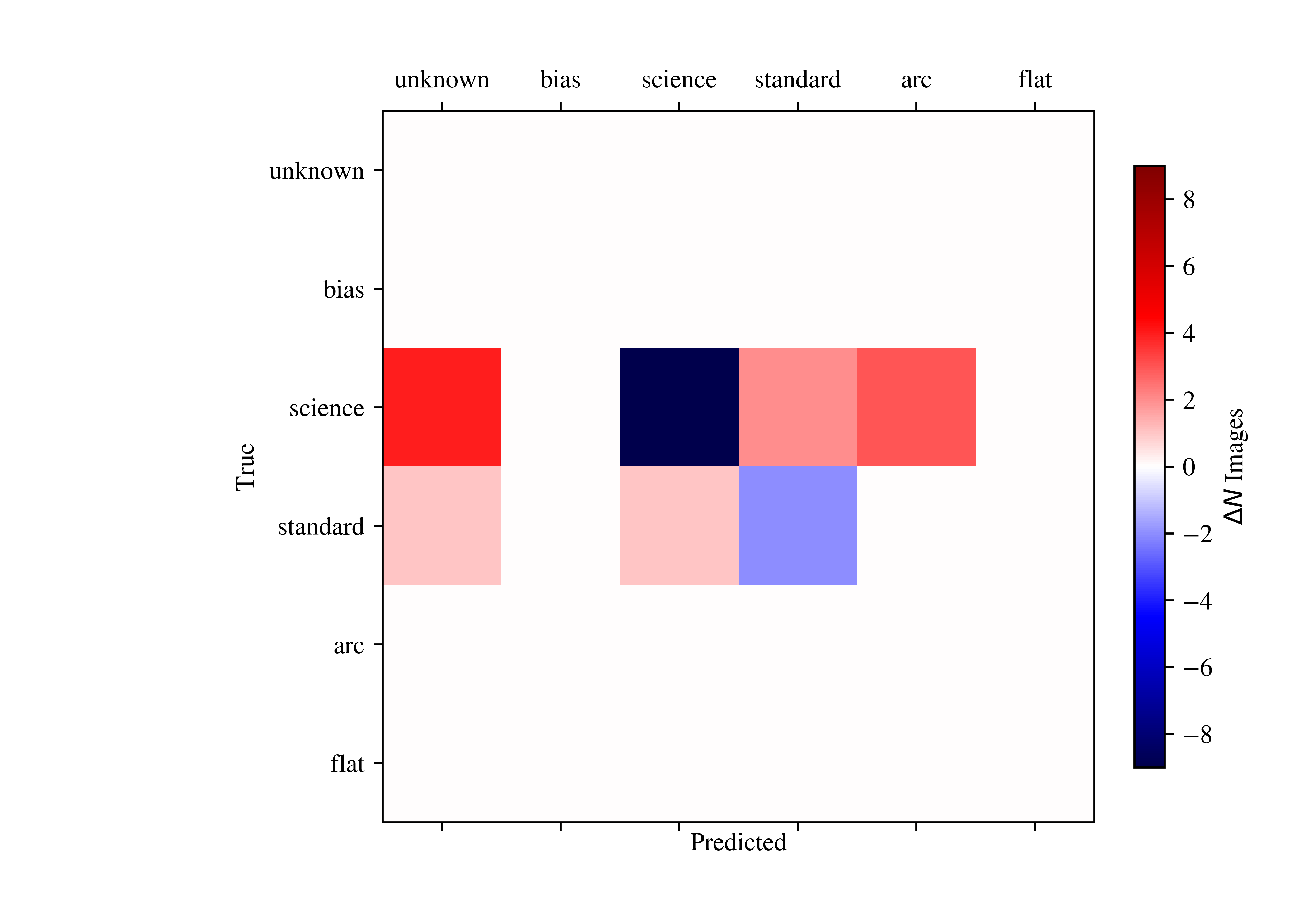}
\caption{Differential $\Delta N$ confusion matrix describing the accuracy
of the optimized CNN classifier on the test suite (unnormalized)
with our final schema.
On the diagonal, negative values indicate the number of images
that were misclassified.  Off the diagonal, the positive values
indicate how the confusion between image types.
A perfect classifier would present a differential confusion
matrix of all zeros.
}
\label{fig:test_final}
\end{center}
\end{figure}

\subsection{Final Accuracy}

After some additional experimentation, we settled on a 
modified classification scheme for any unique image:
(i)  generate one one-hot encoded array for the input image and 
its three flipped counterparts;
(ii) take the majority value of each one-hot encoded array;
(iii) adopt the majority value of these 4 inputs provided
it is a `super'-majority, i.e.\ occurring 3 or 4 times;
(iv) otherwise record the classification as Unknown.
This approach increases the robustness of {\sc SPIT}
and better identifies images that are either corrupt or
especially difficult to classify based on the image data
alone.

With this final schema applied to the 626~test images (i.e.\ without
normalization), we find  \percftest\ are accurately classified,
0.9\%\ were misclassified, and 0.8\%\ were classified as Unknown.
Figure~\ref{fig:test_final} shows the differential confusion 
matrix for this final assessment.

\subsection{Failures}

As described above,
we find that \percftest\ of the test images were correctly
typed, with 100\%\ accuracy for the bias, arc, 
and flat image types.
Our most significant set of failures are in 
distinguishing between Standard and Science
frames.  This was expected: an inexperienced human
would be challenged to distinguish between these
and even an experienced observer could be confused.
Both may show a bright continuum trace and several
sky lines.  The key distinguishing characteristics
are:
  (1) much fewer cosmic rays; 
  (2) fainter sky-line features for the Standard
  frames; 
  and
  (3) brighter sky continuum in the Standard frame.
These are due to the shorter exposure time for
Standard images\footnote{In principle, Science frames 
may have short exposure times too.} and the fact
that standard stars are frequently observed during twilight
when the sky background from scattered sunlight is bright.
We also note several mis-classifications of Science
frames as Arc images.  These are cases where no continuum
was evident (i.e.\ very faint objects) and the classifier
presumably confused the sky emission lines to be arc lines.

\section{Future Work}
\label{sec:future}

Our first implementation of a CNN for spectral image typing 
(\spit) has proven to be highly successful.
Indeed, we are now implementing this methodology in
the \pypit\ DRP beginning with Kast.
Future versions of \spit\ will include training with images
from LRIS, GMOS, HIRES, and other optical/near-IR instruments.
Scientists across the country are welcome and encouraged
to provide additional human-classified images for training
(contact JXP).

As we expand the list to include multi-slit and echelle data,
we may train a more complex CNN architecture.
We will maintain all \spit\ architectures online:
see https://spectral-image-typing.readthedocs.io/en/latest/
for details.


\acknowledgments

We acknowledge the guidance and intellectual 
contributions of D. Parks.  We thank T. Hsyu
for kindly sharing her pre-classified Kast data.
We acknowledge B. Robertson for making available
his GPU computer.
We thank Google for making public their Tensorflow package.

%



\software{Tensorflow \citep{tensorflow2015-whitepaper},
          Adam \citep{Adam2015},
          }

\end{document}